\documentclass[final,3p]{elsarticle}

\usepackage{amsmath,amssymb,amsfonts}
\usepackage{bm}
\usepackage{graphicx}
\usepackage{cleveref}

\journal{Journal of Computational Physics}
\biboptions{sort&compress}

\newcommand{\bu}{\bm{u}}

\newcommand{\pd}[1]{\partial_{#1}}
\newcommand{\dd}[3]{\frac{\text{d}^{#3}#1}{\text{d}#2^{#3}}}
\newcommand{\tJ}{\text{J}}
\newcommand{\sN}{\mathcal{N}}
\newcommand{\mat}[1]{\text{\textbf{\textsf{#1}}}}
\newcommand{\imag}{\text{i}}

\begin{document}

\begin{frontmatter}

\title{Energy-conserving Galerkin approximations for quasigeostrophic dynamics}

\author[CU]{Matthew Watwood}
\author[CU]{Ian Grooms\corref{mycorrespondingauthor}}
\ead{ian.grooms@colorado.edu}
\cortext[mycorrespondingauthor]{Corresponding author}
\author[CU]{Keith A. Julien}
\author[NYU]{K.~Shafer Smith}

\address[CU]{Department of Applied Mathematics, University of Colorado, Boulder, CO 80309}
\address[NYU]{Center for Atmosphere Ocean Science, Courant Institute of Mathematical Sciences, New York University, New York, NY 10012}

\begin{abstract}
A method is presented for constructing energy-conserving Galerkin approximations in the vertical coordinate of the full quasigeostrophic model with active surface buoyancy. 
The derivation generalizes the approach of Rocha \emph{et al.} (2016) \cite{RYG16} to allow for general bases.
Details are then presented for a specific set of bases: Legendre polynomials for potential vorticity and a recombined Legendre basis from Shen (1994) \cite{Shen94} for the streamfunction.
The method is tested in the context of linear baroclinic instability calculations, where it is compared to the standard second-order finite-difference method and to a Chebyshev collocation method.
The Galerkin scheme is quite accurate even for a small number of degrees of freedom $\sN$, and growth rates converge much more quickly with increasing $\sN$ for the Galerkin scheme than for the finite-difference scheme.
The Galerkin scheme is at least as accurate as finite differences and can in some cases achieve the same accuracy as the finite difference scheme with ten times fewer degrees of freedom.
The energy-conserving Galerkin scheme is of comparable accuracy to the Chebyshev collocation scheme in most linear stability calculations, but not in the Eady problem where the Chebyshev scheme is significantly more accurate.
Finally the three methods are compared in the context of a simplified version of the nonlinear equations: the two-surface model with zero potential vorticity.
The Chebyshev scheme is the most accurate, followed by the Galerkin scheme and then the finite difference scheme.
All three methods conserve energy with similar accuracy, despite not having any a priori guarantee of energy conservation for the Chebyshev scheme.
Further nonlinear tests with non-zero potential vorticity to assess the merits of the methods will be performed in a future work.  
\end{abstract}

\begin{keyword}
Quasigeostrophic \sep Spectral \sep Galerkin \sep Legendre
\end{keyword}

\end{frontmatter}

%
%
\section{Introduction}
The well-known quasigeostrophic (QG) model describes the dynamics of extratropical oceanic and atmospheric circulations \cite{Majda03,Pedlosky87,Vallis17}, characterized by the dominant roles of background stratification, leading to hydrostatic balance, and planetary rotation, resulting in geostrophic balance. In the case of a fluid bounded above and below by flat rigid surfaces, where the vertical velocity must vanish, the dynamics are governed by the evolution of three quantities: the quasigeostrophic potential vorticity (PV) $q(x,y,z,t)$ and the buoyancy at the top and bottom surfaces, $b^+(x,y,t)$ and $b^-(x,y,t)$ respectively.
(The superscripts $^+$ and $^-$ henceforth denote evaluation of a quantity at the top and bottom surfaces of the domain, respectively.)
The three quantities $q$ and $b^\pm$ evolve according to 

\begin{subequations}\label{eqn:QGPV}
\begin{equation}\label{eqn:bplus}
\pd{t}b^+ + \bu^+\cdot\nabla b^+ = 0
\end{equation}
\begin{equation}
\pd{t}q + \bu\cdot\nabla q +\beta v= 0\label{eqn:q}
\end{equation}
\begin{equation}
\pd{t}b^- + \bu^-\cdot\nabla b^- = 0.\label{eqn:bminus}
\end{equation}
\end{subequations}
The equations are set in a linear tangent plane to the sphere at latitude $\theta$.
The parameter $\beta$ is $(f/R)\cos(\theta)$ where $R$ is the radius of the Earth and $f$ is twice the rotation rate of the Earth.
Hereafter the symbol $\nabla$ is to be understood as horizontal only.
The velocity is also horizontal, $\bu\cdot\nabla = u\pd{x}+v\pd{y}$, and the velocity is incompressible $\nabla\cdot\bu=0$.
The velocity is obtained from the PV and surface buoyancies by solving the following elliptic equation for the streamfunction $\psi$

\begin{subequations}\label{eqn:Inversion}
\begin{equation}
f_0\pd{z}\psi = b^+\text{ at }z=H
\end{equation}
\begin{equation}
\nabla^2\psi + \pd{z}\left(S(z)\pd{z}\psi\right)=q
\end{equation}
\begin{equation}
f_0\pd{z}\psi = b^-\text{ at }z=0
\end{equation}
\end{subequations}
and then setting $u=-\pd{y}\psi$, $v=\pd{x}\psi$.
The streamfunction $\psi$ can either be set to $0$ on the side boundaries, or they can be periodic.
For periodic boundaries the spatially-constant part of the streamfunction is not determined by (\ref{eqn:Inversion}), but this component of $\psi$ has no impact on the dynamics and can therefore be set to any desired value.
The function $S(z)$ is $f_0^2/N^2(z)$ where $f_0$ is the local Coriolis parameter $f_0 = f\sin(\theta)$ and $N(z)>0$ is the Brunt-V\"ais\"al\"a frequency, also known as the buoyancy frequency.

There are two common simplifications of this system that set either $q$ and $\beta$, or $b^\pm$ to zero.
Both of these simplifications are exact solutions of the full equations.
Setting $b^\pm=0$ leads to considerable simplifications in the analysis of the system and in the development of numerical methods.
For example, solutions of the system with $b^\pm=0$ are known to be regular and to be a regular asymptotic limit of the Navier-Stokes equations \cite{BB94}, and standard finite-difference approximations are known to be convergent \cite{Colin97}.

The other simplification sets $q=\beta=0$.
The surface-QG model (sQG) is obtained by further setting $b^+=0$ and replacing the boundary condition $\pd{z}\psi=0$ at $z=H$ with the condition $\pd{z}\psi\to0$ as $z\to\infty$.
Rigorous mathematical analysis of the inviscid sQG model is considerably more difficult than the model with $b^\pm=0$. 
A connection of the sQG model to the three-dimensional Euler equations was made in \cite{CMT94}, and the study of well-posedness of the sQG model is ongoing (e.g.~\cite{CN17}; for related results in a model with dissipation see \cite{Lapeyre17,NV18}).
As a result of this connection between the full unsimplified QG model and the sQG model global regularity of the full QG system remains an open problem, though strong solutions are known to exist and be unique for a finite time horizon \cite{BK82}.

Surface buoyancy has a significant impact on the dynamics of the upper ocean \cite{LH06} and on the atmosphere near the tropopause \cite{TS09b}, and the simplified system with $b^\pm=0$ is unable to model these dynamics.
Surface-QG is able to model the impact of surface buoyancy on atmospheric and oceanic dynamics, but it is only an exact solution of the full system (\ref{eqn:QGPV}) when $\beta=0$.
In addition, several studies have used sQG theory to infer ocean subsurface velocities using only surface buoyancy but these studies have found that the assumption $q=0$ prevents accurate reconstruction except near the ocean surface \cite{LM06,ILKCH08,KLRLS11}.
To model the interplay of surface and interior dynamics one needs the full unsimplified QG model.\\

The primary difficulty in constructing discretizations of the full QG model is the discretization of the vertical coordinate $z$.
The equations do not have a particularly unusual form and any one of a variety of classical methods could be used, but particular attention is paid in the community to whether a discretization is energy-conserving.
This is because simulations are often used to study the energetics of the system, e.g.~the transfer of energy between horizontal and vertical scales \cite{SV01,SV02,RMCM12}, using integrations over long time scales.
The classical second order finite difference discretization found in \cite{Pedlosky87,Colin97,Vallis17} conserves energy and is the standard method for simulations of the full system \cite{SV01,SV02,HDKB03,KBG09,MMC10,VVS11,RMCM12,SB15,GN16}.
Higher-order alternatives are therefore desirable for the purpose of achieving equal accuracy with less cost, or higher accuracy at equal cost.

For a fully discrete system to be energy-conserving naturally depends also on the discretization of the horizontal directions.
Energy-conserving horizontal discretizations are available \cite{Arakawa66}; our focus is on discretization of the vertical coordinate.

The main alternative to the standard second order finite difference discretization of the vertical coordinate is a Galerkin approach based on \cite{Flierl78}.
The operator $\pd{z}(S(z)\pd{z}\mathbf{\cdot})$ with homogeneous Neumann boundary conditions admits a set of eigenfunctions called `baroclinic modes' that form an orthogonal basis for $L^2(0,1)$.
A finite number of these modes can be used as a basis for an energy-conserving Galerkin approximation of the QG equations, but the straightforward application clearly assumes that $b^\pm=0$ since that is the boundary condition satisfied by the basis functions.
It was recently shown in \cite{RYG16} how the same basis could be used in a way that does not require $b^\pm=0$, and that still conserves energy.
This approach is counter-intuitive in that it generates an approximate solution with $\pd{z}\psi=0$ at the surfaces, but with $b^\pm\neq0$; nevertheless the approximation to $\psi$ still converges absolutely and uniformly as the number of basis functions increases.
Unfortunately these basis functions are not practically useful except in the case where $S(z)$ is a constant, in which case the modes are just Fourier modes.
(Precisely this Fourier baroclinic mode basis was recently used in \cite{SPM18}, but for the simplified QG model with $b^\pm=0$.)

It was proposed in \cite{TS09b} to simply augment the baroclinic mode basis with auxiliary functions that enable satisfaction of the inhomogeneous Neumann conditions, but this method does not conserve energy \cite{TS09b,RYG16}.
An alternative orthogonal basis was developed in \cite{SV13}.
This basis enables both satisfaction of the inhomogeneous Neumann boundary conditions and conservation of energy in a Galerkin approximation, but the basis does not enable separation of variables in the solution of the elliptic equation and is more useful for analysis of observational data than for high-resolution simulations of the nonlinear dynamics.\\

This article presents an energy-conserving Galerkin approximation scheme for the vertical coordinate of the full QG system that generalizes the approach in \cite{RYG16} so that it can be used with any appropriate basis while allowing active surface buoyancy.
We immediately specialize to a global polynomial basis based on Legendre polynomials.
Legendre polynomials are a convenient basis because energy is defined using an un-weighted $L^2$ norm squared, and the Legendre polynomials are orthogonal with respect to the un-weighted $L^2$ inner product. 
The paper is organized as follows. 
Our main result on the construction of an energy-conserving Galerkin approximation is found in \cref{sec:Galerkin}.
Implementation details and a specific choice of polynomial basis are presented in \cref{sec:Legendre}.
The method is tested and compared to the standard finite diffence method and to a non-energy-conserving Chebyshev collocation method in the context of linear baroclinic instability calculations in \cref{sec:Stability}.
In \cref{sec:2SQG} the new Galerkin method is compared to the standard finite difference method and to the Chebyshev collocation scheme in fully-nonlinear simulations of idealized two-surface dynamics with $\beta=q=0$.
Results are discussed and conclusions are offered in \cref{sec:Conclusions}.

%
%
\section{Energy-Conserving Galerkin Approximations}
\label{sec:Galerkin}
The QG equations conserve energy in the form

\[E = \frac{1}{2}\int_\Omega |\nabla\psi|^2+S(z)(\pd{z}\psi)^2\]
where $\int_\Omega$ represents an integral over the whole domain $(x,y,z)\in \Omega$.
The first term corresponds to kinetic energy and the second to available potential energy.
The proof of energy conservation is straightforward: \Cref{eqn:q} is multiplied by $-\psi$, followed by integration over the volume.
Careful use of integration by parts together with \cref{eqn:Inversion} and the surface buoyancy equations (\ref{eqn:bplus}) and (\ref{eqn:bminus}) yield the desired result.
The boundary conditions at the side boundaries can be assumed to be either impenetrable or periodic.

Suppose that $q$ will be represented as a linear combination of basis functions $p_n^q(z)$

\begin{equation}\label{eqn:qsN}
q_\sN = \sum_{n=1}^\sN\breve q_n(x,y,t)p_n^q(z).
\end{equation}
The notation $\sN$ serves to distinguish the number of basis functions $\sN$ from the Brunt-V\"ais\"al\"a frequency $N$.
The notation for the coefficients $\breve q_n$ follows \cite{RYG16}.
Similarly, suppose that $\psi$ will be represented as a linear combination of basis functions $p_n^\psi(z)$, which can be different from the basis used for $q$:

\begin{equation} \label{eqn:psisN}
\psi_\sN = \sum_{n=1}^\sN\breve \psi_n(x,y,t)p_n^\psi(z).
\end{equation}
The approximation to $\psi$ is used to evolve surface buoyancy as follows

\[\pd{t}b^{\pm} + \bu_\sN^\pm\cdot\nabla b^{\pm} = 0.\]

There are now two approximation problems.
The first is related to PV inversion: given $q_\sN$ and $b^\pm$, how does one obtain coefficients $\breve\psi_n$ for $\psi_\sN$?
The second is related to PV evolution: given $q_\sN$ and $\psi_\sN$, how does one obtain the tendencies $\pd{t}\breve q_n$?
The way that these questions are answered determines the kind of approximation being made, as well as whether the scheme conserves energy.

However these questions are answered one can always define residuals $r_q$ and $r_t$ related to the two approximations, along with residuals $r_b^\pm$ related to the boundary conditions

\begin{subequations}
\begin{equation}
r_q = q_\sN - \nabla^2\psi_\sN - \pd{z}\left(S(z)\pd{z}\psi_\sN\right)\label{eqn:InversionG}
\end{equation}
\begin{equation}
r_t = \pd{t}q_\sN +\bu_\sN\cdot\nabla q_\sN + \beta v_\sN\label{eqn:qG}
\end{equation}
\begin{equation}
r_b^\pm = b^\pm - f_0\pd{z}\psi_\sN^\pm.\label{eqn:bG}
\end{equation}
\end{subequations}
It is important to note that in the above equations $q_\sN$ and $\psi_\sN$ are not Galerkin coefficients (which are denoted $\breve q_n$ and $\breve \psi_n$); they are instead the full Galerkin approximations to $q$ and $\psi$ given by (\ref{eqn:qsN}) and (\ref{eqn:psisN}).
If the boundary conditions on the PV inversion are exactly satisfied then $r_b^\pm=0$.
The approximate potential vorticity $q_\sN$ is not materially conserved unless $r_t=0$; this is true regardless of the basis functions chosen, including finite elements or the baroclinic modes of \cite{RYG16}.
A Galerkin approximation to the PV inversion would choose the coefficients $\breve\psi_n$ according to the condition that the residual $r_q$ be orthogonal to the span of the basis functions $p_n^\psi$.

An energy-conserving discretization should conserve energy in the following form

\[E_\sN = \frac{1}{2}\int_\Omega |\nabla\psi_\sN|^2+S(z)(\pd{z}\psi_\sN)^2.\]
One can obtain an exact evolution equation for $E_\sN$ by, for example, multiplying \cref{eqn:qG} by $-\psi_\sN$ and integrating over the volume.
One eliminates $q_\sN$ using \cref{eqn:InversionG}, and then performs integrations by parts using boundary conditions \cref{eqn:bG}.
The result is

\begin{equation}\label{eqn:EnergyG}
\dd{E_\sN}{t}{}=-\int_x\left[S(H)\psi_\sN^+\pd{t}r_b^+-S(0)\psi_\sN^-\pd{t}r_b^-\right]+ \int_\Omega\psi_\sN(\pd{t}r_q-r_{t}).
\end{equation}
The first integral on the right hand side of this equation is taken over the horizontal upper and lower surfaces; this has been indicated by the subscript $x$ on the integral: $\int_x$.

Energy conservation can evidently be achieved quite simply as follows.
First apply the usual Galerkin condition to the PV inversion by requiring $r_q$ to be $L^2$-orthogonal to the basis functions $p_n^\psi(z)$; this eliminates the term $\int_\Omega\psi_\sN\pd{t}r_q$ in \cref{eqn:EnergyG}.
Next, satisfy the boundary conditions exactly so that $r_b^\pm=0$, eliminating the first term on the right hand side of \cref{eqn:EnergyG}.
Finally, apply a Petrov-Galerkin condition to determine the evolution $\pd{t}q_\sN$ by requiring $r_t$ to be $L^2$-orthogonal to the span of $p_n^\psi$, eliminating the last term on the right hand side of \cref{eqn:EnergyG}.
(This latter is a Petrov-Galerkin condition because the residual is made orthogonal to a different subspace than the one in which the approximation is sought. If $p_n^q=p_n^\psi$ then this is just another Galerkin condition.)

The problem is that there are only $2\sN$ degrees of freedom --- one each for $\breve\psi_n$ and $\breve q_n$ --- while the above recipe yields $2\sN+2$ conditions.
This difficulty was avoided in \cite{RYG16} by making use of a clever reformulation of the PV inversion proposed by Bretherton \cite{Bretherton66}.
The solution to the PV inversion \cref{eqn:Inversion} is the same as the solution to the following reformulated problem \cite{Bretherton66}

\begin{subequations}\label{eqn:Bretherton}
\begin{equation}
\pd{z}\psi =0\text{ at }z=H
\end{equation}
\begin{equation}
\nabla^2\psi + \pd{z}\left(S(z)\pd{z}\psi\right)=q - \frac{f_0}{N^2(z)}b^+\delta(z-H) + \frac{f_0}{N^2(z)}b^-\delta(z)
\end{equation}
\begin{equation}
\pd{z}\psi = 0\text{ at }z=0.
\end{equation}
\end{subequations}
The equivalence of these two formulations can be obtained through the Green's function formulation of the solution, as shown in \ref{sec:AppC}.
There is a new residual associated with this reformulation of the inversion, defined to be 

\begin{equation}
r_q^B = q_\sN -\frac{f_0}{N^2(z)}b^+\delta(z-H) + \frac{f_0}{N^2(z)}b^-\delta(z)- \nabla^2\psi_\sN - \pd{z}\left(S(z)\pd{z}\psi_\sN\right).\label{eqn:InversionB}
\end{equation}
The superscript $B$ in $r_q^B$ stands for `Bretherton.'
Using this reformulated problem to derive an evolution equation for the discretized energy yields a deceptively similar equation:

\begin{equation}\label{eqn:Energy0}
\dd{E_\sN}{t}{}=-\int_x\left[S(H)\psi_\sN^+\pd{t}r_b^+-S(0)\psi_\sN^-\pd{t}r_b^-\right]+ \int_\Omega\psi_\sN(\pd{t}r_q^B-r_{t}).
\end{equation}
Further simplifications are possible though, since the basis functions can now be assumed to satisfy homogeneous Neumann boundary conditions $\pd{z}p_n^\psi=0$, consistent with the boundary conditions of the reformulated PV inversion.
These boundary conditions imply that 

\begin{equation}
r_b^\pm = b^\pm\quad\Rightarrow\quad\pd{t}r_b^\pm = -\bu_\sN^\pm\cdot \nabla b^\pm.
\end{equation}
This implies that the first integral on the right hand side of the energy budget \cref{eqn:Energy0} is zero since $\int_x\psi_\sN\bu_\sN\cdot\nabla b=0$.
As a consequence the energy budget takes the form

\begin{equation}\label{eqn:Energy1}
\dd{E_\sN}{t}{}= \int_\Omega\psi_\sN(\pd{t}r_q^B-r_{t}).
\end{equation}
Energy conservation can now be achieved through the use of a Galerkin condition on the reformulated PV inversion ($r_q^B$ $L^2$-orthogonal to $p_n^\psi$), and a Petrov-Galerkin condition on the evolution tendency ($r_t$ $L^2$-orthogonal to $p_n^\psi$).
This is essentially a re-derivation of the result in \cite{RYG16} using an arbitrary basis instead of the baroclinic mode basis.
This derivation enables the use of practical algorithms based on finite element bases, spline bases, or polynomial bases in cases where the baroclinic mode basis of \cite{RYG16} is unavailable or unwieldy.

We note as a brief aside that the Galerkin condition on the PV inversion is equivalent to choosing $\psi_\sN$ to minimize a semi-norm of the error.
To wit, let $\psi_*$ be the true solution to the reformulated PV inversion with $q=q_\sN$ and note that in general $\psi_*$ cannot be exactly described by an approximation of the form \cref{eqn:psisN}; then the $\psi_\sN$ that minimizes the following semi-norm of the error

\begin{equation}
\|\psi_*-\psi_\sN\|_q^2 = \int|\nabla(\psi_*-\psi_\sN)|^2 + S(z)\left(\pd{z}(\psi_*-\psi_\sN)\right)^2
\end{equation}
is the same as the $\psi_\sN$ that sets the residual $r_q^B$ $L^2$-orthogonal to the span of the basis functions $p_n^\psi$.
This was not precisely clear in \cite{RYG16} where the discussion could be misconstrued to suggest that the Galerkin condition is equivalent to minimizing the $L^2$ norm of the error.\\

Energy is not the only sign-definite quadratic quantity conserved by the full quasigeostrophic system (\ref{eqn:QGPV}); it also conserves enstrophy when $\beta=0$.
Enstrophy is half the volume integral of the square of the potential vorticity; in the approximation (\ref{eqn:qsN}) it is 

\[Z_\sN = \frac{1}{2}\int_\Omega q_\sN^2.\]
Its evolution in the approximate system is derived by multiplying \cref{eqn:qG} by $q_\sN$ and integrating over the volume.
The result, assuming $\beta=0$, is

\begin{equation}
\frac{\text{d}Z_\sN}{\text{d}t} = \int_\Omega q_\sN r_t.
\end{equation}
The energy-conserving method proposed above imposes the condition that $r_t$ be $L^2$-orthogonal to the span of the basis functions $p_n^\psi$.
The above expression indicates that approximate enstrophy will not be conserved in the energy-conserving scheme unless the same basis functions are used for both $\psi$ and $q$, i.e. $p_n^\psi = p_n^q$.
The energy-conserving method requires $\pd{z}p_n^\psi = 0$ at the boundaries, but there is no a priori reason to impose the boundary condition $\pd{z}q=0$ by using $p_n^\psi=p_n^q$.
One is therefore faced with the choice of using $p_n^\psi=p_n^q$, which leads to an enstrophy-conserving method but potentially degrades the approximation of $q$, or using $p_n^\psi\neq p_n^q$, which will not conserve enstrophy but may allow a more accurate approximation of $q$.
In the following section we continue the analysis under the more general assumption that $p_n^\psi$ is not necessarily equal to $p_n^q$, leaving open the option to choose $p_n^\psi=p_n^q$ if desired.

%
%
\section{A Legendre basis}
\label{sec:Legendre}
This section begins with general considerations associated with implementation using an arbitrary basis, and continues to consideration of a specific basis using Legendre polynomials.
There are two problems to deal with: (i) computing $\psi_\sN$ from $q_\sN$ and $b^\pm$, and (ii) computing $\pd{t}q_\sN$.

First consider the implementation of the Galerkin condition on the reformulated PV inversion \cref{eqn:Bretherton}.
Assume that the basis $p_n^\psi(z)$ satisfies homogeneous Neumann conditions.
The condition that the residual in \cref{eqn:InversionB} be $L^2$-orthogonal to the basis functions $p_n^\psi$ leads to an $\sN\times\sN$ linear system of the following form

\begin{equation}\label{eqn:DiscretePV}
\nabla^2\mat{M}\bm{\psi} -\mat{L}\bm{\psi} = \mat{B}\mathbf{q} - \frac{f_0}{N^2(H)}b^+\mathbf{p}^{\,+} + \frac{f_0}{N^2(0)}b^-\mathbf{p}^{\,-}
\end{equation}
where the vectors $\bm{\psi}$ and $\mathbf{q}$ contain the Galerkin coefficients $\breve\psi_n$ and $\breve q_n$, the vectors $\mathbf{p}^{\,+}$ and $\mathbf{p}^{\,-}$ have elements

\begin{equation}\label{eqn:PVec}
\mathbf{p}_n^{\,+}=p_n^\psi(H)\qquad\mathbf{p}_n^{\,-}=p_n^\psi(0),
\end{equation}
and the matrices have the following elements

\begin{subequations}
\begin{equation}
\mat{M}_{i,j} = \int_0^Hp_i^\psi(z)p_j^\psi(z)\text{d}z,
\end{equation}
\begin{equation}
\mat{L}_{i,j} = \int_0^HS(z)(\pd{z}p_i^\psi(z))(\pd{z}p_j^\psi(z))\text{d}z,\end{equation}
\begin{equation}
\mat{B}_{i,j} = \int_0^Hp_i^\psi(z)p_j^q(z)\text{d}z.
\end{equation}
\end{subequations}

The discrete PV inversion system \cref{eqn:DiscretePV} can be diagonalized by a change of basis. 
The mass matrix \textbf{\textsf{M}} is symmetric positive definite and has a Cholesky factorization 

\[\mat{M} = \mat{G}\mat{G}^T.\]
The matrix $\mat{G}^{-1}\mat{L}\mat{G}^{-T}$ is symmetric positive semi-definite, and has an orthogonal eigenvalue decomposition

\[\mat{G}^{-1}\mat{L}\mat{G}^{-T} = \mat{Q}\bm{\Sigma}\mat{Q}^T\]
where $\mat{Q}$ is real orthogonal and $\bm{\Sigma}$ is diagonal.
With these matrix decompositions the discrete PV inversion can be diagonalized as follows

\begin{equation}
\nabla^2\mat{M}\bm{\psi} -\mat{L}\bm{\psi} = \mat{GQ}(\nabla^2\mat{I} - \bm{\Sigma})\mat{Q}^T\mat{G}^T\bm{\psi}.	
\end{equation}
The PV inversion has to be solved repeatedly during a time integration of the PV equations.
An efficient implementation first computes the matrices $\mat{G}$, $\mat{Q}$, and $\bm{\Sigma}$, then uses the diagonalization above to split the PV inversion into a set of independent two-dimensional elliptic inversions.
Once the matrix factorizations have been computed, the cost to invert the PV is $\mathcal{O}(\sN^2)$ (plus $\sN$ times the cost of each two-dimensional elliptic inversion).
This approach, of decomposing the three-dimensional PV inversion into a series of independent two-dimensional PV inversions, is employed by several widely-used codes including \cite{SV01,HDKB03,KBG09,Grooms16,GN16}.
The grid sizes in common current use are small enough that several two-dimensional PV inversions fit within local memory in a single node of a distributed-memory machine, which means that cross-node communication is limited to the change of vertical coordinate. 
Furthermore, there are often between several hundred and one thousand degrees of freedom in each horizontal direction, while $\sN$ is usually less than 100 and only occasionally in the low hundreds for codes that use the standard second-order finite difference approximation.
The Galerkin method developed here should require smaller $\sN$ than the finite difference approximation, so in practice $\sN$ should be expected to be relatively small and the $\mathcal{O}(\sN^2)$ cost to convert the three-dimensional inversion to a set of two-dimensional inversions should be significantly smaller than the cost of a single two-dimensional inversion.
Nevertheless, it may be true in some cases that a full multigrid approach applied directly to \cref{eqn:DiscretePV} will be more efficient than the diagonalization described above (cf.~\cite{RMCM12}). 

Each column of the matrix $\mat{G}^{-T}\mat{Q}$ contains coefficients of a function in the basis $p_n^\psi$.
These functions together form a basis that diagonalizes the discrete PV inversion operator.
These basis functions that diagonalize the discrete PV operator are approximations to the baroclinic modes of \cite{RYG16} that diagonalize the continuous PV operator.\\

Next consider the problem of computing $\pd{t}q_\sN$ via the Petrov-Galerkin condition that $r_t$ in \cref{eqn:qG} be $L^2$-orthogonal to the $p_n^\psi$ basis.
This condition yields a system of the form

\begin{equation}\label{eqn:Bq}
\mat{B}\left(\pd{t}\mathbf{q}\right) = \dot{\mathbf{q}}
\end{equation}
where the elements of the right hand side are

\begin{equation}
\left(\dot{\mathbf{q}}\right)_n =-\int_0^Hp_n^\psi(\bu_\sN\cdot\nabla q_\sN + \beta v_\sN)\text{d}z.
\end{equation}
Fortunately, for basis functions consisting of polynomials or piecewise polynomials these integrals can be computed exactly, either analytically or via appropriate quadratures.
Since this system needs to be solved repeatedly during a time integration, an LU decomposition of the matrix {\bf B} can be computed once before starting the integration.
In the specific basis developed in the next section, the matrix $\mat{B}$ has nonzero entries only on the diagonal and on the second super-diagonal.
The cost to solve \cref{eqn:Bq} in this case is $3\sN-4$ floating point operations.

\subsection{A Legendre basis}
The foregoing analysis applies to any set of basis functions $p_n^q$ and $p_n^\psi$ with the assumption $\pd{z}p_n^\psi=0$ at the boundaries.
Note that as in \cite{RYG16} this does not imply that $b^\pm=0$.
This section considers a specific choice of the two basis sets.
First, since orthogonality is of necessity defined using the $L^2$ inner product, it is convenient to let $p_n^q(z) = L_{n-1}(z)$ where $L_k(z)$ is the $k^\text{th}$ order Legendre polynomial, rescaled to the interval $z\in[0,H]$.

Legendre polynomials do not satisfy $\pd{z}L_k(z)=0$ at the boundaries, and therefore cannot be used for $p_n^\psi$.
Shen \cite{Shen94} constructs the following functions

\begin{equation}\label{eqn:Shen}
\phi_k(z) = L_k(z) - \frac{k(k+1)}{(k+2)(k+3)}L_{k+2}(z).
\end{equation}
These functions form a basis for polynomials with homogeneous Neumann conditions, and we set $p_n^\psi(z) = \phi_{k-1}(z)$.

\begin{figure}[t]
\centering
\includegraphics[width=.5\textwidth]{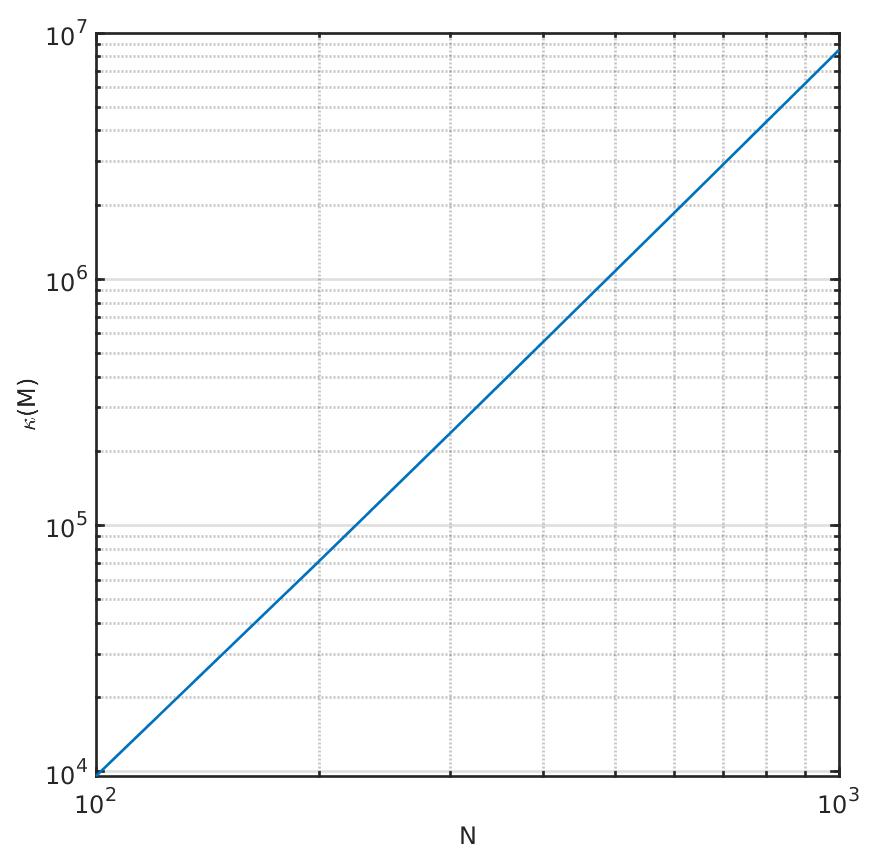}
\caption{\label{fig:CondM} The two-norm condition number of the mass matrix $\mat{M}$ for $\sN$ from 100 to 1000.}
\end{figure}

With this choice of $p_n^q$ and $p_n^\psi$ the $\mat{B}$ matrix is upper triangular with upper bandwidth 2.
The mass matrix {\bf M} is pentadiagonal; in fact, an even-odd permutation converts the $\mat{M}$ matrix to a block-diagonal matrix with tridiagonal blocks.
The elements of the matrices $\mat{M}$ and $\mat{B}$ are known analytically.
The two-norm condition number of $\mat{M}$ is plotted in \cref{fig:CondM} for $\sN$ from 100 to 1000.
The condition number is just under $10^7$ for $\sN=1000$; a condition number of $10^7$ is moderate for a banded symmetric positive definite matrix in double precision arithmetic, and $\sN=1000$ is much larger than would be used in most applications.
For example, the maximum $\sN$ used in the finite-difference simulations of \cite{MMC10,RMCM12} was 128. 

The structure of the $\mat{L}$ matrix depends on the stratification $S(z)$ and is in general dense.
(For a finite element basis the $\mat{L}$ matrix would be sparse.)
The elements of $\mat{L}$ can in general be computed using quadrature.
In the computations described in \cref{sec:Stability} the elements of $\mat{L}$ are computed using Gauss-Legendre quadrature.\\

With $\sN$ Legendre basis functions $q_\sN$ is a polynomial in $z$ of degree $\le \sN-1$, while $\psi_\sN$ (and hence $\bu_\sN$) is a polynomial in $z$ of degree $\le \sN+1$.
The elements of the vector $\dot{\mathbf{q}}$ are therefore integrals of polynomials of degree at most $3\sN+1$.
These can be evaluated exactly (up to roundoff error) using Gauss-Legendre quadrature with $1.5\sN+1$ nodes.

%
%
\section{Linear Baroclinic Instability}
\label{sec:Stability}

This section makes a preliminary assessment of the accuracy of the energy-conserving Legendre Galerkin scheme described in the preceding section by applying it to the linear quasigeostrophic baroclinic instability problem.
Any configuration of the form 

\[\bar{\psi} = -y\bar{u}(z),\;\;\bar{q}(z) = -y\dd{}{z}{}\left(S(z)\dd{\bar{u}}{z}{}\right),\;\;\bar{b}^+ = -yf_0\left.\dd{\bar{u}}{z}{}\right|_{z=H},\;\;\bar{b}^- = -yf_0\left.\dd{\bar{u}}{z}{}\right|_{z=0}\]
is an exact solution of the fully nonlinear QG equations \cref{eqn:QGPV}.
The linearization of the QG equations about an equilibrium of this form is

\begin{subequations}
\begin{equation}
\pd{t}b^++\bar{u}^+\pd{x}b^+ - \left(f_0\dd{\bar u}{z}{}\right)_{z=H}v^+ = 0
\end{equation}
\begin{equation}
\pd{t}q+\bar{u}\pd{x}q +(\pd{y}\bar{q}+\beta)v = 0
\end{equation}
\begin{equation}
\pd{t}b^-+\bar{u}^-\pd{x}b^- - \left(f_0\dd{\bar u}{z}{}\right)_{z=0}v^- = 0.
\end{equation}
\end{subequations}
Linear instability about an equilibrium of this form is called `baroclinic instability.'
The linear equations vary only in $z$, so it is convenient to Fourier transform in the horizontal direction, and to assume exponential growth in time with the form $e^{-\imag k_x c t}$ where $k_x$ is the wavenumber in the $x$ direction and $c$ is the wave phase speed.
Also replacing $f_0\text{d}\bar u/\text{d}z$ by $-\pd{y}\bar{b}$ on the boundaries leads to equations of the form

\begin{subequations}\label{eqn:ExactLinear}
\begin{equation}\label{eqn:ExactLinearBPlus}
\bar{u}^+\hat{b}^+ + (\pd{y}\bar{b}^+)\hat{\psi}^+ = c\hat{b}^+
\end{equation}
\begin{equation}\label{eqn:ExactLinearQ}
\bar{u}\hat{q} +(\pd{y}\bar{q}+\beta)\hat{\psi} = c\hat{q}
\end{equation}
\begin{equation}\label{eqn:ExactLinearBMinus}
\bar{u}^-\hat{b}^- + (\pd{y}\bar{b}^-)\hat{\psi}^- = c\hat{b}^-
\end{equation}
\end{subequations}
where $\hat{q}$ is the Fourier transform of $q$ and $\hat{b}^\pm$ is the Fourier transform of $b^\pm$.
The standard energy-conserving second-order finite difference discretization of the linear stability problem is described in \ref{sec:AppB}.
We also use a standard, non-energy-conserving Chebyshev collocation method to discretize the linear stability problem; Chebyshev collocation methods for linear problems are described in a variety of places including \cite{Trefethen00,CHQZ06}.

\subsection{Galerkin Discrete Linear Baroclinic Instability}
To discretize this system (\cref{eqn:ExactLinear}) according to the methods described in the previous section, one makes Galerkin approximations to both the equilibrium state and the perturbations, then one imposes Galerkin and Petrov-Galerkin conditions on the residuals.
First consider how to construct the appropriate Galerkin approximation to the equilibrium state.
Although the equilibrium can be completely described by $\bar u(z)$, the correct approach within the method described in the foregoing section is to first produce a Galerkin approximation to $\pd{y}\bar q$ and then invert to find $\bar u$.
The expansion coefficients in the approximation of $\pd{y}\bar q$ are arranged into the vector $\mathbf{\bar q}_y$, whose entries are

\[\left(\mathbf{\bar q}_y\right)_n = \frac{\int_0^Hp_n^q(z)(\pd{y}\bar q)\text{d}z}{\int_0^H(p_n^q(z))^2\text{d}z}.\]
(This expression assumes that the basis functions $p_n^q(z)$ are $L^2$-orthogonal, which is true for the Legendre basis considered here.
For a non-orthogonal basis one would have to solve a linear system to find $\mathbf{\bar q}_y$.)
Once this vector is available, the coefficients in the Galerkin approximation of $\bar u$ are obtained by solving the following system

\begin{equation}\label{eqn:BarUInversion}
\mat{L}\mathbf{\bar u} = -\mat{B}\mathbf{\bar q}_y+\frac{f_0}{N^2(H)}(\pd{y}\bar b^+)\mathbf{p}^{\,+} - \frac{f_0}{N^2(0)}(\pd{y}\bar b^-)\mathbf{p}^{\,-}.
\end{equation}
The vectors $\mathbf{p}^\pm$ are defined in \cref{eqn:PVec}.

Unfortunately the matrix $\mat{L}$ is singular: its first row and column are zero because $p_1^\psi(z)=1$ and $\pd{z}p_1^\psi=0$.
Fortunately the right hand side is always compatible: the first entry of the right hand side is always zero as well.
This statement is substantiated in \ref{sec:AppA}.
The first entry of $\mathbf{\bar u}$ is not constrained by the linear system above; it corresponds to the depth-independent component of $\bar u(z)$.
The first Galerkin coefficient of $\bar u$ can simply be set to

\[\mathbf{\bar u}_1 = \frac{\int_0^Hp_1^\psi(z)\bar u\text{d}z}{\int_0^H(p_1^\psi(z))^2\text{d}z}.\]
The remaining entries of $\mathbf{\bar u}$ are obtained by solving the lower-right $\sN-1\times\sN-1$ block of \cref{eqn:BarUInversion}.\\

Next consider how to construct an appropriate Galerkin approximation to the perturbations about the equilibrium.
The Galerkin approximations to the perturbations $\hat{q}$ and $\hat{\psi}$ will have coefficients stored in the vectors $\mathbf{q}$ and $\bm{\psi}$, respectively.
These coefficients are related through the Fourier transform of \cref{eqn:DiscretePV}, which is

\begin{equation}\label{eqn:FourierDiscretePV}
-(k_x^2+k_y^2)\mat{M}\bm{\psi} -\mat{L}\bm{\psi} = \mat{B}\mathbf{q} - \frac{f_0}{N^2(H)}b^+\mathbf{p}^{\,+} + \frac{f_0}{N^2(0)}b^-\mathbf{p}^{\,-}.
\end{equation}

With Galerkin approximations to the equilibrium and perturbations in hand, one next inserts these approximations into \cref{eqn:ExactLinearQ} and requires the residual to be orthogonal to $p_n^\psi$.
One also inserts $\bar{u}_\sN$ in place of $\bar u$ in \cref{eqn:ExactLinearBPlus} and \cref{eqn:ExactLinearBMinus}, yielding the following linear system

\begin{subequations}\label{eqn:LinearUnsimplified}
\begin{equation}
\bar{u}_\sN^+\hat{b}^+ +(\pd{y}\bar{b}^+)\hat{\psi}_\sN^+ = c\hat{b}^+
\end{equation}
\begin{equation}
\overline{\mat{U}}\mathbf{q} +(\overline{\mat{Q}}_y+\beta\mat{M})\bm{\psi} = c\mat{B}\mathbf{q}
\end{equation}
\begin{equation}
\bar{u}_\sN^-\hat{b}^- +(\pd{y}\bar{b}^-)\hat{\psi}_\sN^- = c\hat{b}^-.
\end{equation}
\end{subequations}
Note that $\hat{\psi}_\sN^\pm = \bm{\psi}\cdot\mathbf{p}^\pm$, where the vectors $\mathbf{p}^\pm$ are defined in \cref{eqn:PVec}.
With this notation the linear system \cref{eqn:LinearUnsimplified} can be written

\begin{subequations}
\begin{equation}
\bar{u}_\sN^+\hat{b}^+ +(\pd{y}\bar{b}^+)\mathbf{p}^{\,+}\cdot\bm{\psi} = c\hat{b}^+
\end{equation}
\begin{equation}
\overline{\mat{U}}\mathbf{q} +(\overline{\mat{Q}}_y+\beta\mat{M})\bm{\psi} = c\mat{B}\mathbf{q}
\end{equation}
\begin{equation}
\bar{u}_\sN^-\hat{b}^- +(\pd{y}\bar{b}^-)\mathbf{p}^{\,-}\cdot\bm{\psi} = c\hat{b}^-.
\end{equation}
\end{subequations}
The matrices $\overline{\mat{U}}$ and $\overline{\mat{Q}}_y$ have the following entries

\begin{equation}
\overline{\mat{U}}_{ij}=\int_0^Hp_i^\psi(z)p_j^q(z)\bar{u}_\sN(z)\text{d}z
\end{equation}
\begin{equation}
\left(\overline{\mat{Q}}_y\right)_{ij}=\int_0^Hp_i^\psi(z)p_j^\psi(z)(\pd{y}\bar{q}_\sN(z))\text{d}z.
\end{equation}

To obtain an eigenvalue problem, the vector $\bm{\psi}$ can be eliminated using \cref{eqn:FourierDiscretePV}

\begin{equation}
\bm{\psi} = -\left((k_x^2+k_y^2)\mat{M}+\mat{L}\right)^{-1}\mat{B}\mathbf{q} + \frac{f_0}{N^2(H)}b^+\bm{\psi}^+ - \frac{f_0}{N^2(0)}b^-\bm{\psi}^-.
\end{equation}
For notational convenience the following vectors have been defined

\begin{equation}
\bm{\psi}^\pm=\left((k_x^2+k_y^2)\mat{M}+\mat{L}\right)^{-1}\mathbf{p}^\pm.
\end{equation}
The matrix $(k_x^2+k_y^2)\mat{M}+\mat{L}$ is invertible as long as $k_x^2+k_y^2\neq0$ since $\mat{M}$ is positive definite and $\mat{L}$ is positive semi-definite.

Eliminating $\bm{\psi}$ leads to a generalized eigenvalue problem of the form

\begin{equation}
\left[\begin{array}{c|c|c}
a_{11}&\mathbf{a}_{12}^T&a_{13}\\
\hline
\mathbf{a}_{21}&\mat{A}_{22}&\mathbf{a}_{23}\\
\hline
a_{31}&\mathbf{a}_{32}^T&a_{33}\end{array}\right]\left(
\begin{array}{c}\hat{b}^+\\\mathbf{q}\\\hat{b}^-\end{array}\right) =
c\left[\begin{array}{c|c|c}
1&\mathbf{0}^T&0\\
\hline
\mathbf{0}&\mat{B}&\mathbf{0}\\
\hline
0&\mathbf{0}^T&0\end{array}\right]\left(
\begin{array}{c}\hat{b}^+\\\mathbf{q}\\\hat{b}^-\end{array}\right)
\end{equation}
where the matrix on the left hand side has the following entries

\begin{subequations}
\begin{equation}
a_{11} = \bar{u}_\sN^++\frac{f_0}{N^2(H)}(\pd{y}\bar{b}^+)\mathbf{p}^{\,+}\cdot\bm{\psi}^+
\end{equation}
\begin{equation}
\mathbf{a}_{12} = -\frac{f_0}{N^2(H)}(\pd{y}\bar{b}^+)\mat{B}^T\left((k_x^2+k_y^2)\mat{M}+\mat{L}\right)^{-1}\mathbf{p}^{\,+}
\end{equation}
\begin{equation}
a_{13} = -\frac{f_0}{N^2(H)}(\pd{y}\bar{b}^+)\mathbf{p}^{\,+}\cdot\bm{\psi}^-
\end{equation}
\begin{equation}
\mathbf{a}_{21} = (\overline{\mat{Q}}_y+\beta\mat{M})\bm{\psi}^+
\end{equation}
\begin{equation}
\mat{A}_{22} = \overline{\mat{U}} -(\overline{\mat{Q}}_y+\beta\mat{M})\left((k_x^2+k_y^2)\mat{M}+\mat{L}\right)^{-1}\mat{B}
\end{equation}
\begin{equation}
\mathbf{a}_{23} = -(\overline{\mat{Q}}_y+\beta\mat{M})\bm{\psi}^-
\end{equation}
\begin{equation}
a_{31} = \frac{f_0}{N^2(0)}(\pd{y}\bar{b}^-)\mathbf{p}^{\,-}\cdot\bm{\psi}^+
\end{equation}
\begin{equation}
\mathbf{a}_{32} = -\frac{f_0}{N^2(0)}(\pd{y}\bar{b}^-)\mat{B}^T\left((k_x^2+k_y^2)\mat{M}+\mat{L}\right)^{-1}\mathbf{p}^{\,-}
\end{equation}
\begin{equation}
a_{33} = \bar{u}_\sN^--\frac{f_0}{N^2(0)}(\pd{y}\bar{b}^{\,-})\mathbf{p}^{\,-}\cdot\bm{\psi}^-.
\end{equation}
\end{subequations}
To solve a specific linear baroclinic instability problem, one chooses external parameters $f_0$, $\beta$, and $N^2(z)$, and an equilibrium state $\bar u(z)$.
Then one computes eigenvalues of the generalized eigenvalue problem, typically over some range of values of $k_x$ and $k_y$; eigenvalues $c$ with positive imaginary part are associated with linearly unstable solutions.
Code to set up and solve the Galerkin and finite-difference linear stability problems is available in \cite{Code}.

\subsection{The Eady Problem}
\begin{figure}[t]
\centering
\includegraphics[width=\textwidth]{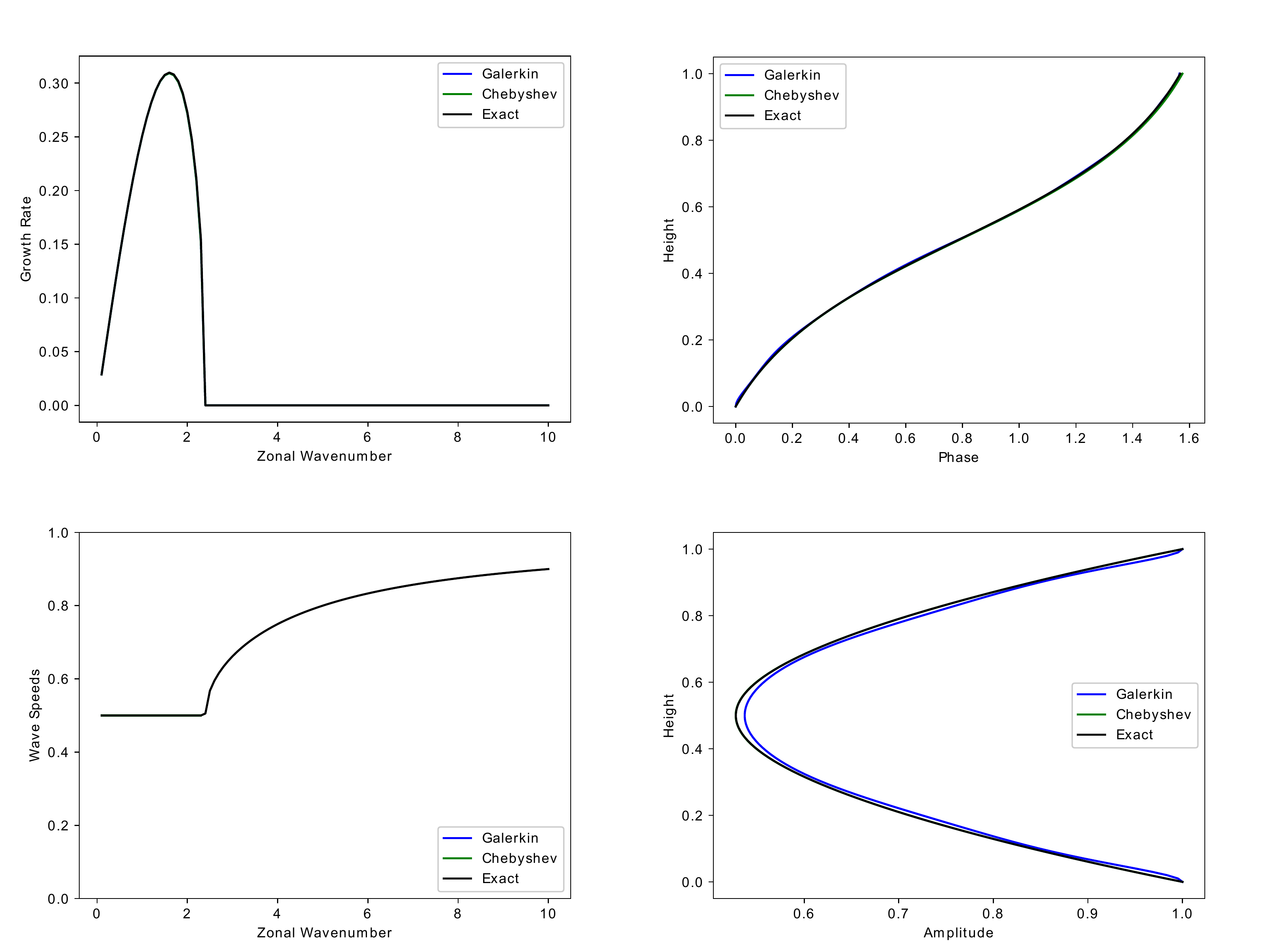}
\caption{\label{fig:Eady1} Comparing Galerkin (blue) and Chebyshev (green) with $\sN=7$  to exact results (black) in the Eady Problem with $k_y=0$. Upper left: Growth rates as a function of $k_x$. Lower left: Wave speeds (real part of the eigenvalue $c$) as a function of $k_x$. Upper right: Complex phase as a function of $z$ for the eigenfunction associated with the fastest-growing mode. Lower right: Amplitude as a function of $z$ for the eigenfunction associated with the fastest-growing mode, normalized to $1$ at $z=0$. The Galerkin and Chebyshev methods are so accurate that the results are indistinguishable from exact in the above plots, except for the Galerkin method in the lower-right panel.}
\end{figure}

The classical Eady problem is defined by constant $N^2(z)$, linear velocity $\bar{u}(z)$, and $\beta=0$.
In this case the baroclinic modes of \cite{RYG16} are simply Fourier modes and are tractable analytically and computationally.
The exact linear perturbation equations \cref{eqn:ExactLinear} are also analytically solvable in the Eady problem (see, e.g.~\cite[Chapter 6]{Vallis17}), which makes for a good test problem.
This section sets $N^2(z) = f_0^2=1$, $H=1$, and $\bar{u} = z$.
The most unstable solutions are found along the axis $k_y=0$, so the generalized eigenvalue problem is solved for a range of $k_x$.

\Cref{fig:Eady1} shows the results of the linear Eady problem with $\sN=7$, compared to the analytical results from \cite[Chapter 6]{Vallis17}.
The growth rates (upper left) and wave speeds (lower left) as a function of $k_x$ are extremely well reproduced with only $\sN=7$ basis functions, to the point where the plots are indistinguishable.
(Note that the streamfunction of the most unstable mode, $k_x\approx 1.6$, is approximated with a polynomial of degree 9.)
The amplitude (upper right) and phase (lower right) of the eigenfunction corresponding to the most unstable mode are also very accurate using only $\sN=7$.
The approximate eigenfunction has $\pd{z}\hat{\psi}=0$ on the boundaries, while the exact does not; nevertheless, the approximate eigenfunction still converges pointwise to the true eigenfunction with increasing $\sN$ (not shown).

\begin{figure}[t]
\centering
\includegraphics[width=.5\textwidth]{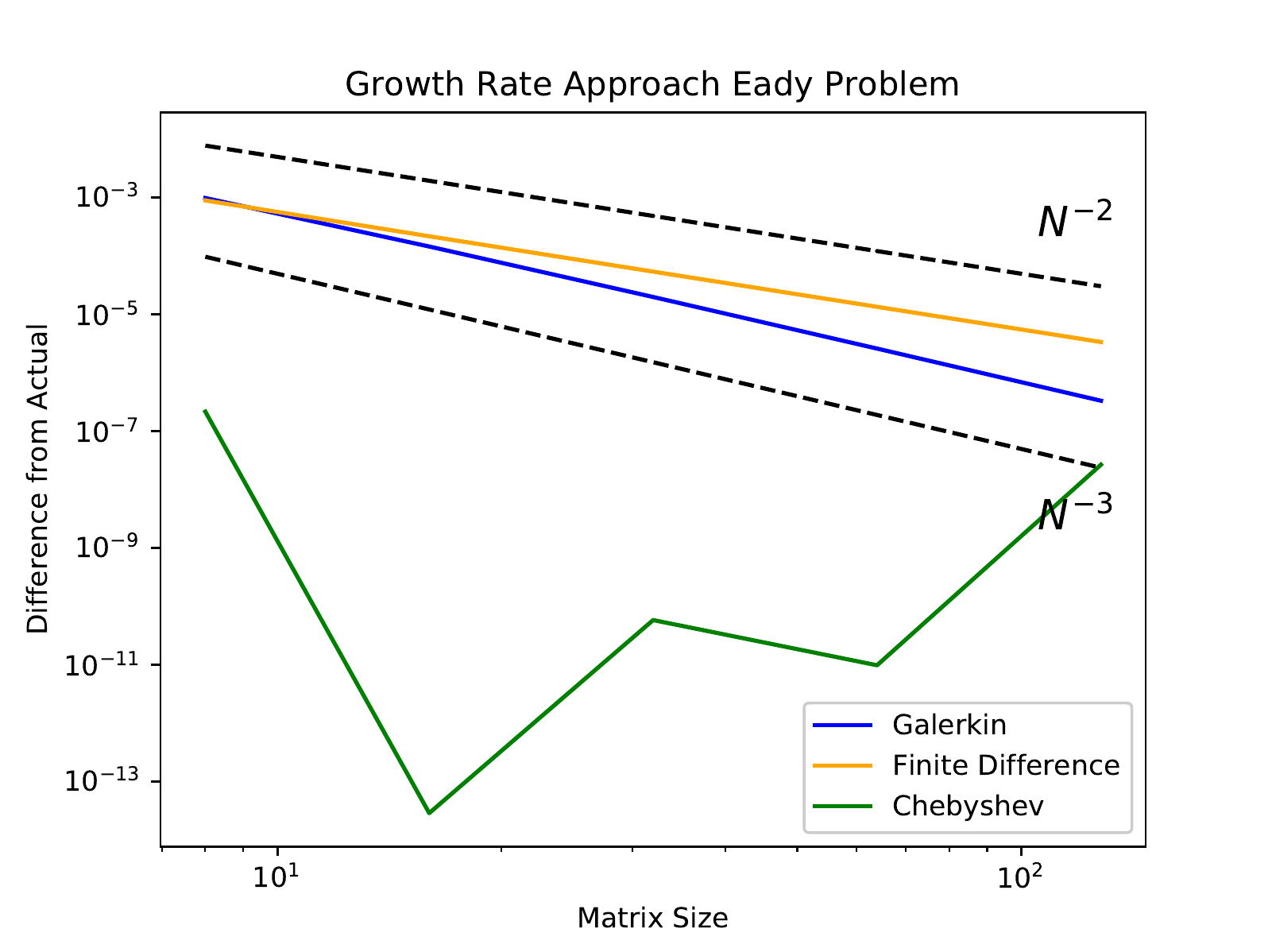}
\caption{\label{fig:Eady2} Error in the growth rate of the fastest growing mode as a function of $\sN$ for the  Galerkin (blue), Chebyshev (green), and finite-difference (orange) methods.}
\end{figure}

\Cref{fig:Eady2} shows the absolute value of the error in the growth rate of the most unstable mode as a function of $\sN$ for the Galerkin approximation, the finite-difference discretization (see \ref{sec:AppB}), and the Chebyshev discretization.
Note that in the finite-difference approximation there are $\sN$ degrees of freedom, while in the Galerkin approximation there are $\sN+2$ degrees of freedom: one for each Galerkin coefficient and one for each surface buoyancy.
The error in the Galerkin approximation decreases as $\mathcal{O}(\sN^{-3})$, while the error in the finite difference approximation decreases only quadratically.
The Chebyshev method exhibits spectral accuracy as expected, with far greater accuracy than the other two methods; accuracy reaches a plateau with increasing $\sN$, which may be a result of the ill-conditioning of Chebyshev differentiation matrices \cite{Fornberg96,Trefethen00,CHQZ06}.
Problems of this sort are avoided by the Chebyshev Galerkin methods of \cite{JW09}.

Convergence of spectral methods can be limited by lack of smoothness in the functions being approximated, but the eigenfunctions of the Eady problem are entire functions expressible as a sum of hyperbolic sine and cosine functions \cite{Vallis17}.
The fact that the Galerkin method converges algebraically rather than exponentially is therefore presumably due to the mismatch between the homogeneous boundary conditions satisfied by the basis functions $p_n^\psi$ and the inhomogeneous boundary conditions satisfied by the true eigenfunctions.
Since this mismatch only occurs in the value of the derivative on the boundary the approximate eigenfunctions still converge to the true eigenfunctions with increasing $\sN$ (as in \cite{RYG16}), though the rate of convergence is slower than it would be if the boundary conditions matched.

\subsection{The Phillips Problem}
\begin{figure}[t]
\centering
\includegraphics[width=\textwidth]{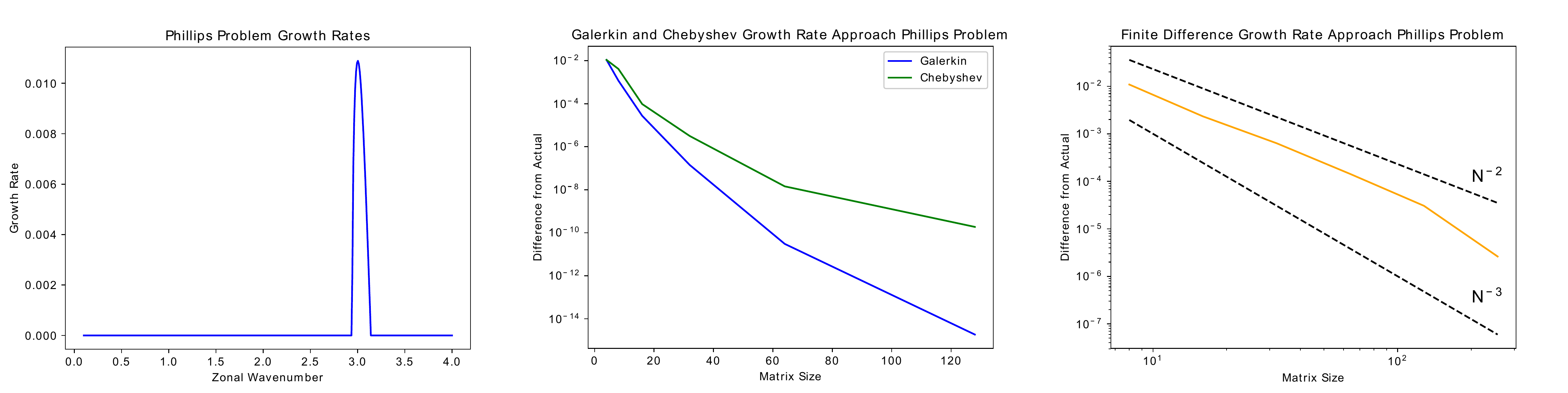}
\caption{\label{fig:Phillips} Growth rates in the Phillips problem. Left: Growth rates versus $k_x$ for the Galerkin method with $\sN=256$. Center: Error in the growth rate at $k_x=3$ as a function of $\sN$ for the Galerkin (blue) and Chebyshev (green) methods. Right: Error in the growth rate at $k_x=3$ as a function of $\sN$ for the finite-difference method.}
\end{figure}

The instability in the Eady problem is driven by interacting edge waves, and is therefore of a type not often seen in the atmosphere or ocean.
Another classical linear baroclinic instability problem that, unlike the Eady problem, is observed in the oceans occurs when the potential vorticity gradient d$\bar q/$d$y$ changes sign in the interior of the fluid.
The canonical representation of this kind of instability is the `Phillips' problem which is distinguished by an equilibrium velocity that has zero shear at the top and bottom surface, and a single sign change in the potential vorticity in the interior.
We construct a `Phillips' problem of this type as follows

\begin{equation}
f_0=N(z)=H=1,\;\;\beta=3.1,\;\;\;\;\bar{u} = -\pi^{-1}\cos(\pi z),\;\;\bar{q} = \pi\cos(\pi z)y.
\end{equation}
The total potential vorticity gradient is $3.1+\pi\cos(\pi z)$.
The negative potential vorticity gradient near the top boundary has small amplitude, and as a result the equilibrium is only slightly above the threshold for instability.
There is only a small range of wavenumbers near $k_x=3$ that are unstable, as shown in the left panel of \cref{fig:Phillips}.

The center and right panels of \cref{fig:Phillips} show the absolute value of the error in the growth rate at $k_x=3$ as a function of $\sN$ for the Galerkin (center), Chebyshev (center), and finite-difference (right) methods.
The center panel uses a logarithmic scale on the growth rate axis and a linear scale on the $\sN$ axis to show that the Galerkin and Chebyshev methods are converging exponentially rather than algebraically.
Although both methods converge exponentially, the Galerkin method is more accurate than the Chebyshev method.
The right panel uses a logarithmic scaling on both axes to show that the finite-difference method is converging quadratically, as usual.
Exponential convergence is expected for the Galerkin method in this case, since the eigenfunctions are smooth and have the same homogeneous Neumann boundary conditions as the basis functions $p_n^\psi$.
Note that the accuracy of the finite difference scheme with $\sN=256$ can be achieved by the Galerkin scheme with ten times fewer degrees of freedom.

\subsection{A Charney-Type Problem}
\begin{figure}[t]
\centering
\includegraphics[width=\textwidth]{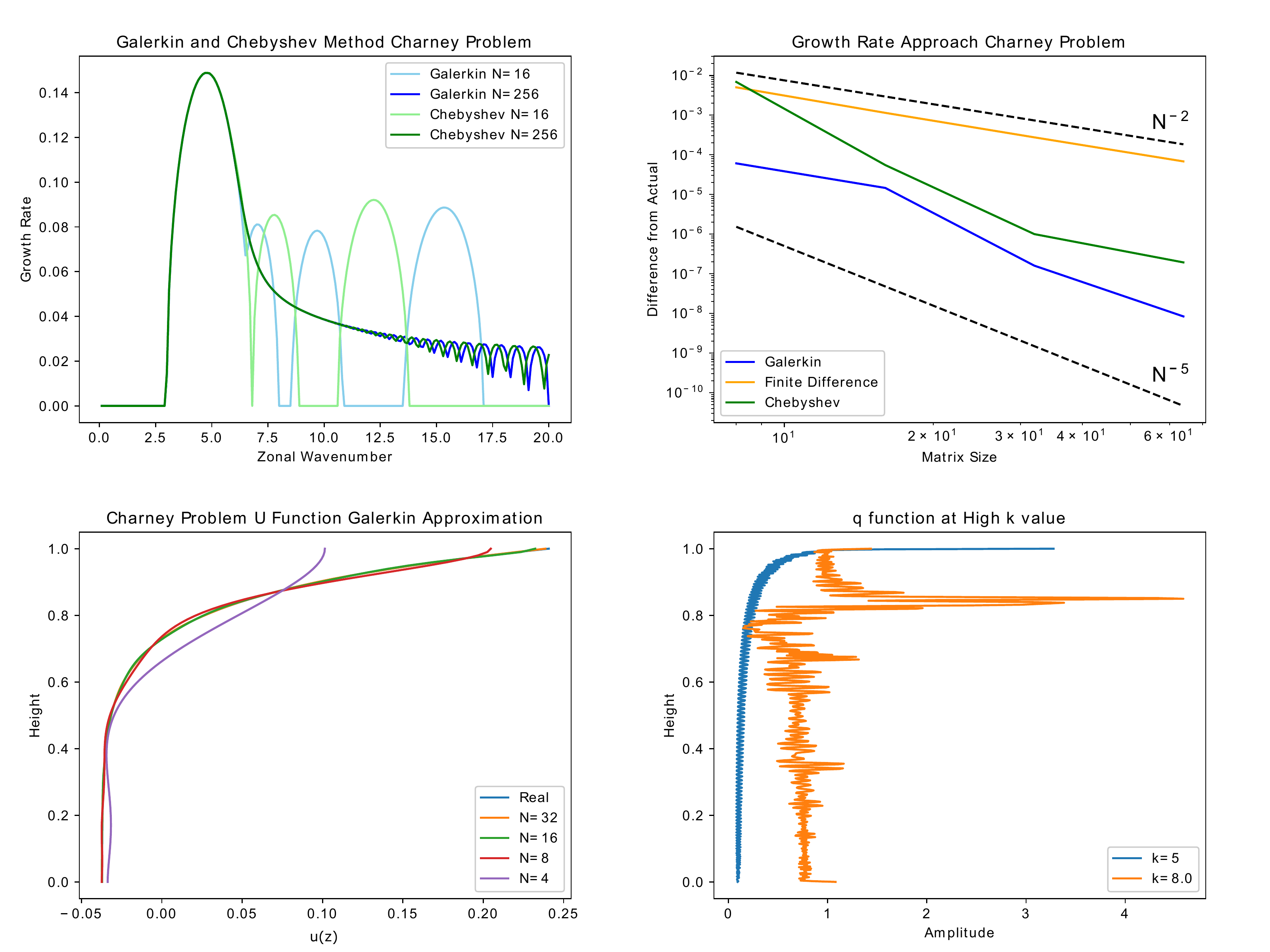}
\caption{\label{fig:OC} Growth rates in the Charney-type problem. Upper left: Growth rates versus $k_x$ for the Galerkin (blue) and Chebyshev (green) methods with $\sN=32$ (pale) and $\sN=256$ (dark). Upper right: Error in the growth rate of the fastest-growing mode as a function of $\sN$ for the Galerkin (blue), Chebyshev (green), and finite-difference (orange) methods. Lower left: Background velocity $\bar{u}(z)$ (blue) and Galerkin approximations $\bar{u}_\sN(z)$ for $\sN=4,\,8,\,16$, and 32. Lower right: Amplitude of the eigenfunction $|\hat{q}(z)|$ at $k_x=5$ (blue) and 8 (orange) using $\sN=256$.}
\end{figure}

Another common type of baroclinic instability in the ocean is driven by the interaction of an edge wave with a Rossby wave in the interior of the fluid \cite{TMHS11}.
A canonical problem describing this kind of instability is the Charney problem, but the canonical Charney problem is posed in a semi-infinite domain with no upper surface.
A Charney-type problem more relevant to the ocean is defined by having a nonzero shear d$\bar u/$d$z$ at the top surface, zero shear at the bottom surface, and a constant potential vorticity gradient d$\bar q/$d$y$ in the interior.
We construct such a Charney-type problem with exponential rather than constant stratification to demonstrate the ability of the Galerkin method to handle non-constant stratification.
The equilibrium state is defined as follows

\begin{equation}
f_0=\beta=H=1,\;\;N^2(z) = e^{6z-6},\;\;\bar{u} = \frac{1}{54}\left(3e^{6z-6}(6z-1)-2-e^{-6}\right),\;\;\bar{q} = -2y.
\end{equation}
Both the stratification and the velocity are surface-intensified (the velocity $\bar{u}(z)$ is shown in the lower right panel of \cref{fig:OC}), which leads to surface-intensification of the unstable linear eigenfunctions.

\Cref{fig:OC} shows the results of the linear Charney-type problem.
The upper left panel of \cref{fig:OC} shows the growth rate as a function of $k_x$ for $\sN=32$ and 256 using the Galerkin method; the finite difference and Chebyshev methods are extremely similar (not shown).
The upper right panel of \cref{fig:OC} shows the absolute value of the error in the growth rate of the most unstable mode as a function of $\sN$ for Galerkin, Chebyshev, and finite-difference methods.
The growth rate in the Galerkin and Chebyshev methods converge approximately quintically ($\mathcal{O}(\sN^{-5})$) while the finite-difference method converges approximately quadratically.
Unlike the Eady problem, the Galerkin approximation is more accurate than both the Chebyshev and the finite difference approximations for the entire range of $\sN$.
The accuracy of the finite difference scheme with $\sN=256$ can be achieved with the Galerkin scheme with ten times fewer degrees of freedom.

A distinctive feature of the Charney-type problem is the presence of weak instability at small scales (large $k_x$), as shown in the upper left panel of \cref{fig:OC}.
Representation of the small-scale instabilities clearly requires large $\sN$; this is true in both the Galerkin and finite-difference methods, which behave similarly for large $k_x$ (not shown).
These small-scale unstable modes result from the interaction of a Rossby wave in a thin layer near the upper surface with an edge wave propagating along the surface.
The lower-right panel of \cref{fig:OC} shows the eigenfunction structure $|\hat{q}_\sN(z)|$ at both $k_x=5$ and $k_x=8$, both computed using $\sN=256$; the near-surface layer is evident in the eigenfunction at $k_x=8$.
(See also fig.~7 of \cite{RYG16}.)
Even at $\sN=256$ these modes are clearly poorly resolved.
As a result of the near-surface nature of the instability, the instability is especially sensitive to the representation of the equilibrium background velocity $\bar{u}(z)$ near the boundary.
The lower left panel of \cref{fig:OC} shows the background velocity $\bar{u}$ along with the Galerkin approximations $\bar{u}_\sN(z)$ for $\sN=4,\,8,\,16$, and 32.
Convergence of $\bar{u}_\sN(z)$  to $\bar{u}(z)$ is slow near the upper boundary because the basis functions satisfy $\pd{z}p_n^\psi=0$ at the boundary, while the equilibrium background profile $\bar{u}(z)$ has $\pd{z}\bar{u}\neq0$ at the boundary.
This slow convergence of $\bar{u}_\sN(z)$ near the boundary is ultimately why the small-scale (high $k_x$) instability shown in the upper left panel of \cref{fig:OC} converges slowly.

%
%
\section{Nonlinear Simulations}
\label{sec:2SQG}
\begin{figure}[t]
\centering
\includegraphics[width=\textwidth]{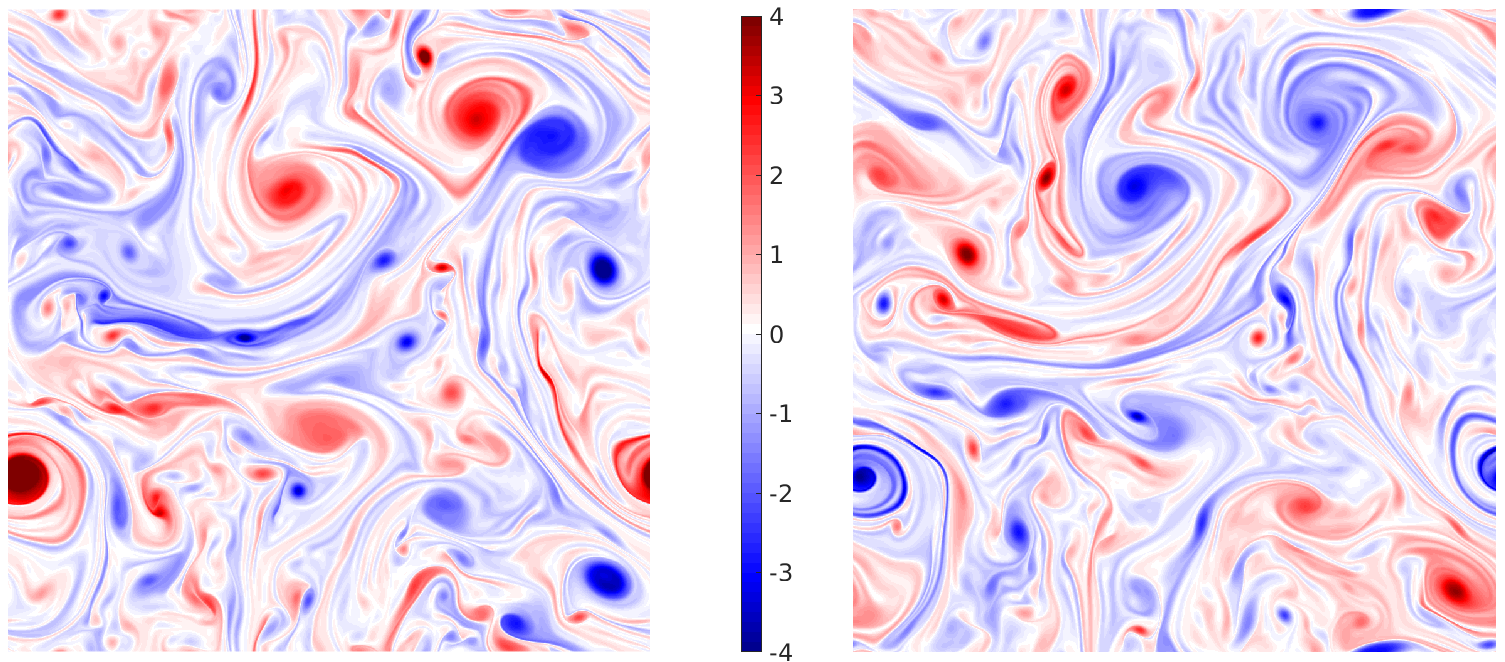}
\caption{\label{fig:2SQG} The state of $b^+$ (left) and $b^-$ (right) used to both estimate inversion accuracy and to initialize simulations tracking energy changes.}
\end{figure}
This section makes a preliminary assessment of the accuracy of the energy-conserving Galerkin scheme described in \cref{sec:Legendre} by using it in fully-nonlinear simulations.
Assessment of the method in comparison with other competing methods using simulations of the full system \cref{eqn:QGPV} will be postponed to a future work; this section uses a simplified exact solution of the full system with $q=\beta=0$.
We further specify $f_0 = N = H = 1$, which leads to the following system of two-dimensional partial differential equations

\begin{subequations}\label{eqn:2SQG}
\begin{equation}\label{eqn:2Sbplus}
\pd{t}b^+ + \bu^+\cdot\nabla b^+ = 0,\;\;\pd{z}\psi = b^+\text{ at }z=1
\end{equation}
\begin{equation}
\left(\nabla^2 + \pd{z}^2\right)\psi=0\label{eqn:2Sq}
\end{equation}
\begin{equation}
\pd{t}b^- + \bu^-\cdot\nabla b^- = 0,\;\;\pd{z}\psi=b^-\text{ at }z=0.\label{eqn:2Sbminus}
\end{equation}
\end{subequations}
This system is very close to the nonlinear Eady model \cite{TS09a} except that it lacks both a mean shear and dissipation terms.
This simplified version of the nonlinear Eady model is used here because it conserves energy exactly, unlike the full nonlinear Eady model, and is therefore an apt test case for comparing our energy-conserving method.
The inversion from $b^\pm$ to $\psi^\pm$ can be solved analytically by means of a Fourier transform

\begin{equation}
\left(\begin{array}{c}\hat{\psi}^+\\\hat{\psi}^-\end{array}\right) = \frac{1}{k}
\left[\begin{array}{rr}
\text{coth}(k)&-\text{csch}(k)\\
\text{csch}(k)&-\text{coth}(k)\end{array}\right]\left(\begin{array}{c}\hat{b}^+\\\hat{b}^-\end{array}\right)
\end{equation}
where $k^2 = k_x^2+k_y^2$.
The three methods used in the preceding section are also used here to solve for $\psi^\pm$.
The Galerkin method uses \cref{eqn:DiscretePV}, the finite difference method is described in \ref{sec:AppB}, and the Chebyshev collocation method is described in several places including \cite{Trefethen00,CHQZ06}.

The domain is a periodic square of width 16$\pi$, and the advection terms in \cref{eqn:2Sbplus} and \cref{eqn:2Sbminus} are discretized using a dealiased Fourier spectral method with $1024$ points in each direction; this method is energy-conserving provided that the vertical discretization is also energy-conserving.
Time integration is achieved via a fourth-order semi-implicit method as described in \cite{GM14}.
This time integration method is not energy-conserving.
Energy conservation typically requires an implicit method, but energy non-conservation due to time integration errors is usually small, and is small in our simulations described below.
For a wide class of systems energy conservation is required in the spatial discretization but not in the temporal discretization in order to achieve time-asymptotic accuracy in the representation of a system's ergodic invariant measure \cite{GTWWW12,Wang16}.
The code used here is a slight adaptation of the publicly-available code \cite{QG_DNS}.\\

\begin{figure}[t]
\centering
\includegraphics[width=\textwidth]{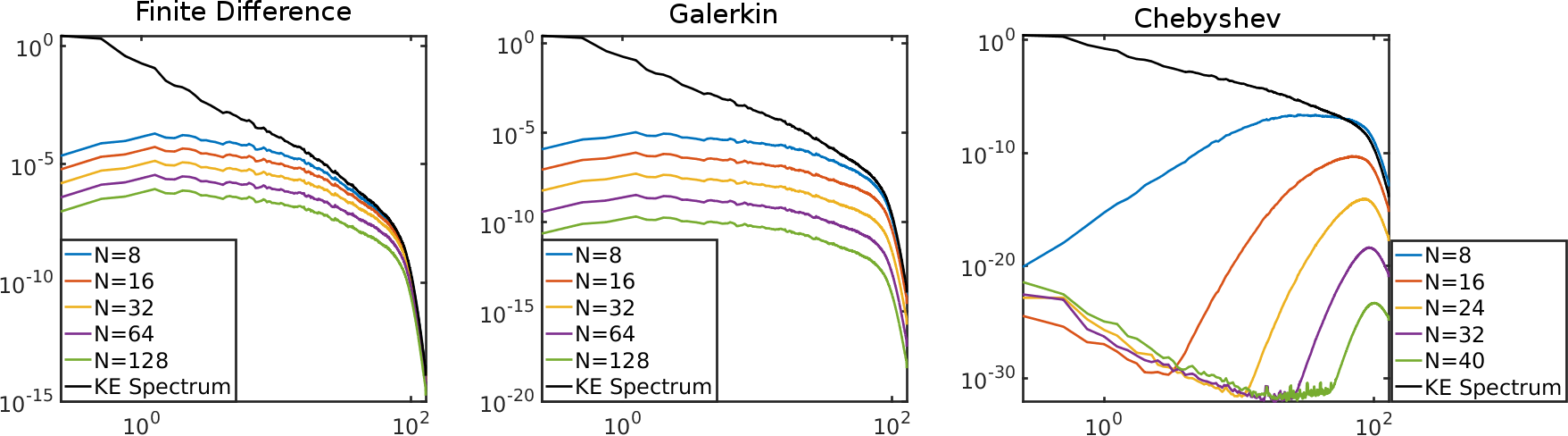}
\caption{\label{fig:ErrSpec} Error kinetic energy spectra at the top surface for the three methods: finite difference (left), Galerkin (center), and Chebyshev (right). Note (i) that the vertical axis scaling is different in each panel, and (ii) that the values of $\sN$ for the Chebyshev method are smaller than for the other two methods. The kinetic energy spectrum is shown in black in each panel for comparison.}
\end{figure}

We begin by spinning up the system from random initial conditions using the exact analytical inversion.
The simulation was stopped when $b^+$ and $b^-$ reached realistic values shown in \cref{fig:2SQG}.
We used this configuration of surface buoyancy to compare the inversion accuracy of the three approximate methods: Galerkin, Chebyshev, and finite-difference.
We computed the value of $\psi^+$ and $\bu^+$ using the exact inversion and the three approximate methods, and then computed the Fourier transform of the error in $\bu^+$.
The one-dimensional spectrum of the velocity error, i.e.~the error kinetic energy spectrum, for each method is shown in \cref{fig:ErrSpec} for different values of $\sN$; for comparison, the true kinetic energy spectrum is also shown.
The Galerkin method (center panel) is clearly more accurate than the finite-difference method (left panel) for a fixed value of $\sN$; the error also decreases faster with increasing $\sN$ for the Galerkin method than for the finite difference method.
On the other hand, the Chebyshev method is so much more accurate than the other two methods that there is essentially no comparison.
This echoes the results from the Eady linear stability problem, which also has constant stratification $S(z)=1$.
In the linear stability problem with non-constant stratification the Chebyshev method was slightly less accurate than the Galerkin method, so the extreme accuracy of the Chebyshev method in \cref{fig:ErrSpec} may not be indicative of its performance generally.

We next assessed the impact of energy conservation by running four simulations from the initial condition shown in \cref{fig:2SQG}: one with the exact inversion and one with each of the three approximate inversions.
These simulations were run for 50 time units (which corresponds to about 40 eddy turnover times), and the total energy was tracked.
The finite difference simulation used 128 levels, the Galerkin simulation used $\sN=16$, and Chebyshev simulations were run with $\sN=4, 8,$ and 16.
In all simulations the energy changed by less than 1\% over the course of the simulation, excepting the Chebyshev simulation with $\sN=4$, which only changed by 1.3\%.
The degree of energy conservation for all three methods was thus near perfect, despite the Chebyshev method not being guaranteed to conserve energy; the observed changes in energy are attributable to the fact that the time integration scheme does not conserve energy.
Further fully nonlinear simulations with non-constant stratification and $q\neq0$ are needed to fully assess the relative merits of the three methods.
These are left for future work.

%
%
\section{Conclusions}
\label{sec:Conclusions}
This article presents an energy-conserving Galerkin approximation scheme for the vertical direction of the full QG system with active surface buoyancy.
The scheme generalizes the Galerkin scheme of \cite{RYG16}.
The method in \cite{RYG16} uses Sturm-Liouville eigenfunctions as a basis to approximate both the potential vorticity $q$ and the streamfunction $\psi$, but these functions are in most cases computationally intractable.
The scheme presented here generalizes the method of \cite{RYG16} to allow an arbitrary basis for $q$ and any basis for $\psi$ that satisfies homogeneous Neumann boundary conditions at the top and bottom surfaces.
Attention is then focused on a Legendre basis for $q$, and a recombined Legendre basis from \cite{Shen94} for $\psi$.
Energy is defined using an unweighted $L^2$ norm, based on an unweighted $L^2$ inner product; Legendre polynomials were used because their orthogonality with respect to the unweighted $L^2$ inner product make them particularly convenient.
Chebyshev polynomials are used more commonly than Legendre polynomials in spectral methods partly because of the existence of a fast transform for Chebyshev polynomials, but Chebyshev polynomials are orthogonal with respect to a weighted $L^2$ inner product and are therefore less convenient than Legendre polynomials in the current setting.

The method was tested and compared to the standard energy-conserving second-order finite-difference method and to a Chebyshev collocation method in the context of linear stability calculations.
In these calculations the Galerkin scheme converged much faster with respect to increasing $\sN$ (vertical resolution) than the finite difference scheme.
The eigenvalues computed with the finite difference scheme converged quadratically, while those computed with the Galerkin scheme converged either at fifth order (in cases with nonzero surface buoyancy) or exponentially (with zero surface buoyancy).
In some cases the Galerkin scheme was able to achieve comparable accuracy to the finite difference method using ten times fewer grid points.
In one case the Chebyshev scheme was very significantly more accurate than the Galerkin scheme, but in two other cases the Galerkin scheme was similar to and slightly more accurate than the Chebyshev method.

The method was then compared to the finite-difference and Chebyshev methods in the context of fully-nonlinear simulations in the simplified setting of $q=\beta=0$, and with constant stratification.
In this simplified setting the only dependent variables are surface buoyancies making the dynamics two-dimensional, though the surface buoyancies are still coupled by a three-dimensional elliptic inversion.
All three methods conserved energy to extremely high accuracy, despite the fact that the Chebyshev scheme is not guaranteed to do so a priori.
The Galerkin method was significantly more accurate than the finite difference method, but was not nearly as accurate as the Chebyshev method.

The scheme presented here is ultimately intended for use in a fully nonlinear, fully three-dimensional setting with nonzero potential vorticity $q$.
The linear stability computations and simplified nonlinear simulations presented here only give limited insight into the accuracy of the scheme for the intended application.
The scheme presented here will be tested in a fully nonlinear and three-dimensional setting in future work.
Finite element bases rather than a global polynomial basis could also be explored as a means of enhancing sparsity in future work.

%
%
\appendix
\section{Compatibility of the system \cref{eqn:BarUInversion}} 
\label{sec:AppA}
The first entry of the right hand side of \cref{eqn:BarUInversion} is a sum of three components, and this appendix demonstrates that the sum of these three components is zero.
The proof relies on the fact that $p_1^\psi(z) = 1$, $p_j^q(z) = L_{j-1}(z)$, and $p_1^q(z)=1$.

The first component of the right hand side of \cref{eqn:BarUInversion} is the first entry of the vector $-\mat{B}\mathbf{\bar q}_y$.
The first row of the matrix {\bf B} has elements

\[\int_0^Hp_1^\psi(z)p_j^q(z)\text{d}z=\int_0^Hp_j^q(z)\text{d}z = H\delta_{1j}\]
where $\delta_{ij}$ is the Kronecker delta.
The first entry of $-\mat{B}\mathbf{\bar q}_y$ is thus simply the first entry of $\mathbf{\bar q}_y$ multiplied by $-H$.

The first entry of $\mathbf{\bar q}_y$ is 

\begin{align*}
\frac{\int_0^Hp_1^q(z)\pd{y}\bar{q}(z)\text{d}z}{\int_0^H(p_1^q(z))^2\text{d}z} &= -\frac{1}{H}\int_0^H\dd{}{z}{}\left(S(z)\dd{\bar u}{z}{}\right)\text{d}z \\
&= \frac{1}{H}\left[S(0)\left.\dd{\bar u}{z}{}\right|_{z=0}-S(H)\left.\dd{\bar u}{z}{}\right|_{z=H}\right]\\
&= \frac{1}{H}\left[\frac{f_0}{N^2(H)}\pd{y}\bar{b}^+-\frac{f_0}{N^2(0)}\pd{y}\bar{b}^-\right].
\end{align*}
Multiplying this by $-H$ results in an expression that exactly cancels the remaining two components on the right hand side of \cref{eqn:BarUInversion}.

\section{Finite Difference Discretization} 
\label{sec:AppB}
This section recalls the standard finite-difference discretization of the QG equations, which can be found in, e.g., \cite{Pedlosky87} and \cite{Vallis17}.
A derivation with careful treatment of surface buoyancy and unequal spacing can be found in \cite{GN16}.

Let $\Delta_z = 1/\sN$ be the grid spacing where $\sN$ is the number of vertical levels.
Both $\psi$ and $q$ are tracked at $\sN$ points starting at $z_1=\Delta_z/2$ and ending at $z_\sN=1-\Delta_z/2$.
The finite difference approximation to $\nabla^2\psi+\pd{z}(S(z)\pd{z}\psi)$ at an interior point $z_k$ ($k\neq1,\sN$) is

\begin{equation}\label{eqn:FDInversionInterior}
\left(\nabla^2\psi+\pd{z}(S(z)\pd{z}\psi)\right)|_{z=z_k} \approx\nabla^2\psi_k+ \frac{1}{\Delta_z}\left[S_k\frac{\psi_{k+1}-\psi_k}{\Delta_z}-S_{k-1}\frac{\psi_k-\psi_{k-1}}{\Delta_z}\right] = q_k
\end{equation}
where $S_k = S(k\Delta_z)$.
At the boundaries we have the following approximations

\begin{gather}
\left(\nabla^2\psi+\pd{z}(S(z)\pd{z}\psi)\right)|_{z=z_1} \approx \nabla^2\psi_1+\frac{1}{\Delta_z}\left[S_1\frac{\psi_{2}-\psi_1}{\Delta_z}-\frac{f_0}{N^2(0)}b^-\right] = q_1\\
\left(\nabla^2\psi+\pd{z}(S(z)\pd{z}\psi)\right)|_{z=z_\sN} \approx \nabla^2\psi_\sN+\frac{1}{\Delta_z}\left[\frac{f_0}{N^2(H)}b^+-S_{\sN-1}\frac{\psi_\sN-\psi_{\sN-1}}{\Delta_z}\right] = q_\sN.
\end{gather}

As discussed in \cite{GN16}, if one defines

\begin{subequations}\label{eqn:FDInversion1N}
\begin{equation}
Q_1 = q_1+\frac{f_0}{\Delta_zN^2(0)}b^- = \nabla^2\psi_1+\frac{1}{\Delta_z}\left[S_1\frac{\psi_{2}-\psi_1}{\Delta_z}\right],
\end{equation}
\begin{equation}
Q_\sN = q_\sN-\frac{f_0}{\Delta_zN^2(H)}b^+ = \nabla^2\psi_\sN-\frac{1}{\Delta_z}\left[S_{\sN-1}\frac{\psi_\sN-\psi_{\sN-1}}{\Delta_z}\right]
\end{equation}
\end{subequations}
Then the fully nonlinear system dynamics are controlled entirely by the following system

\begin{subequations}\label{eqn:FDDynamics}
\begin{equation}
\pd{t}Q_1 + \tJ[\psi_1,Q_1] + \beta\pd{x}\psi_1 = 0
\end{equation}
\begin{equation}
\pd{t}q_k + \tJ[\psi_k,Q_k] + \beta\pd{x}\psi_k = 0,\;\;k = 2,\ldots,\sN-1
\end{equation}
\begin{equation}
\pd{t}Q_\sN + \tJ[\psi_\sN,Q_\sN] + \beta\pd{x}\psi_\sN = 0.
\end{equation}
\end{subequations}
The only caveat is that by evolving this system one knows $Q_1$ and $Q_\sN$ but not $b^\pm$ or $q_1$ and $q_\sN$, but the dynamics of $\psi_k$ are completely controlled by the above system: \cref{eqn:FDDynamics} for the dynamics and \cref{eqn:FDInversionInterior} and \cref{eqn:FDInversion1N} for the PV inversion.\\

The discrete version of the linear stability problem is straightforward in the finite difference approximation.
One can start with \cref{eqn:ExactLinear} and then discretize as described above.
The discrete finite difference problem takes the form of the following generalized eigenvalue problem

\[\left[\overline{\mat{U}}_{FD}\left((k_x^2+k_y^2)\mat{I}+\mat{L}_{FD}\right)-\left(\overline{\mat{Q}}_{y,FD}+ \beta\mat{I}\right)\right]\bm{\psi} = c\left[(k_x^2+k_y^2)\mat{I}+\mat{L}_{FD}\right]\bm{\psi}.\]

The matrix $\overline{\mat{U}}_{FD}$ is diagonal with diagonal elements $\bar u(z_k)$.
The matrix $\mat{L}_{FD}$ is tridiagonal with the form

\[\mat{L}_{FD} = \frac{1}{\Delta_z^2}\left[\begin{array}{ccccc}
S_1&-S_1&0&\cdots&0\\
-S_1&S_1+S_2&-S_2&&\vdots\\
\vdots&\ddots&\ddots&\ddots&\vdots\\
&-S_{k-1}&S_{k-1}+S_k&-S_k&\\
\vdots&\ddots&\ddots&\ddots&\vdots\\
0&\cdots&0&-S_{\sN-1}&S_{\sN-1}
\end{array}\right]\]

The matrix $\overline{\mat{Q}}_{y,FD}$ is also diagonal.
If one defines a vector $\bar{\bm{u}}$ whose elements are $\bar u(z_k)$, the diagonal elements of $\overline{\mat{Q}}_{y,FD}$ are the elements of the vector $\mat{L}_{FD}\bar{\bm{u}}$.
It is interesting to note that in the Galerkin method the approximate velocity profile $\bar{u}_\sN(z)$ is derived from the potential vorticity gradient and the surface bouyancy gradients, while in the finite-difference approximation the potential vorticity gradient is derived from the velocity profile.

\section{Bretherton's Formulation}
\label{sec:AppC}
This section formally demonstrates the equivalence of the original PV inversion problem \cref{eqn:Inversion} with Bretherton's \cite{Bretherton66} reformulation \cref{eqn:InversionB}.
The goal is to show that despite having imposed homogeneous Neumann boundary conditions on the streamfunction in Bretherton's formulation, the presence of Dirac delta distributions on the right hand side ensures that the actual solution satisfies the same inhomogeneous boundary conditions as the original problem.
In the context of the discretized problem, the approximate solution will satisfy homogeneous Neumann boundary conditions on both boundaries for any $\sN$.
The discrete approximation is nevertheless still expected to converge {\it pointwise} to the true solution as $\sN\to\infty$.

We begin by giving the Fourier transform of the Bretherton problem \cref{eqn:InversionB}

\begin{equation}\label{eqn:C1}
    -k^2\hat{\psi} + \pd{z}\left(S(z)\pd{z}\hat{\psi}\right) = \hat{q} - \frac{f_0\hat{b}^+}{N^2(z)}\delta(z-H) + \frac{f_0\hat{b}^-}{N^2(z)}\delta(z),\qquad \pd{z}\hat{\psi}=0\text{ at }z=0,H.
\end{equation}
Away from the boundaries the above equation is exactly the same as the original formulation, so the question of equivalence of the two formulations centers on the behavior of $\pd{z}\hat{\psi}$ on the boundaries.
The Green's function formulation of the solution is

\begin{align}\notag
    \hat{\psi}(z;k) &= \int_0^Hg(z,s;k)\left(\hat{q}(s)- \frac{f_0\hat{b}^+}{N^2(s)}\delta(s-H) + \frac{f_0\hat{b}^-}{N^2(s)}\delta(s)\right)\text{d}s\\
    &=\int_0^Hg(z,s;k)\hat{q}(s)\text{d}s - \frac{f_0\hat{b}^+}{N^2(H)}g(z,H;k) + \frac{f_0\hat{b}^-}{N^2(0)}g(z,0;k)
\end{align}
where $g(z,s;k)$ the the Green's function.
The derivative of $\hat{\psi}$ with respect to $z$ is formally

\begin{equation}\label{eqn:dpsidz}
    \pd{z}\hat{\psi}(z;k) =\int_0^H\pd{z}g(z,s;k)\hat{q}(s)\text{d}s - \frac{f_0\hat{b}^+}{N^2(H)}\pd{z}g(z,H;k) + \frac{f_0\hat{b}^-}{N^2(0)}\pd{z}g(z,0;k).
\end{equation}
The Green's function can be written in the form \cite[Chapter 10]{HN01}

\begin{equation}
    g(z,s;k) = \frac{1}{S(s)(w_1(s)w_2'(s) - w_1'(s)w_2(s))}\left\{\begin{array}{cl}
    w_1(z)w_2(s)&0\le z\le s\\
    w_1(s)w_2(z)&s\le z\le 1\end{array}\right.
\end{equation}
where $w_1$ and $w_2$ are functions satisfying

\begin{subequations}
\begin{equation}
    -k^2w_i(z) + \pd{z}\left(S(z)\pd{z}w_i(z)\right) = 0\text{ for }i=1,2
\end{equation}
\begin{equation}
    w_1'(0) = 0,\quad w_1'(H) = 1
\end{equation}
\begin{equation}
    w_2'(0) = 1,\quad w_2'(H) = 0
\end{equation}
\end{subequations}
and the notation $w_1'$ denotes the derivative of $w_1$.
Consider $\pd{z}g(z,s;k)$ at $z=0$ for any $s\neq0$

\begin{equation}
    [\pd{z}g(z,s;k)]_{z=0} = \frac{w_1'(0)w_2(s)}{S(s)(w_1(s)w_2'(s) - w_1'(s)w_2(s))} = 0.
\end{equation}
A similar manipulation shows that $\pd{z}g(z,s;k)=0$ at $z=H$ for any $s\neq H$.
This shows that the contribution to $\pd{z}\hat{\psi}$ from the integral in \cref{eqn:dpsidz} is zero on the boundaries $z=0,H$.
Now consider $\pd{z}g(z,0;k)$.
At $s=0$ the Green's function is

\begin{equation}
g(z,0;k) = \frac{w_2(z)}{S(0)}
\end{equation}
and the derivative is

\begin{equation}
    \pd{z}g(z,0;k) = \frac{w_2'(z)}{S(0)}.
\end{equation}
At $z=0$ we have $\pd{z}g(z,0;k) = S(0)^{-1} = N^2(0)/f_0^2$, while at $z=H$ we have $\pd{z}g(z,0;k) = 0$.
A similar argument shows that at $z=H$ $\pd{z}g(z,H;k) = -S(H)^{-1} = -N^2(H)/f_0^2$ while at $z=0$ $\pd{z}g(z,H;k) =0$.
Plugging these expressions into \cref{eqn:dpsidz} and evaluating at $z=0,H$ yields

\begin{equation}
    \pd{z}\hat{\psi}|_{z=0} = f_0b^-\text{ and } \pd{z}\hat{\psi}|_{z=H} = f_0b^+.
\end{equation}
(The foregoing argument is a more general version of the proof carried out in appendix A of \cite{TS09a} for constant $N^2(z)$.)
This seems to contradict the homogeneous Neumann boundary conditions imposed at the beginning of this section.
The paradox is explained by realizing that the presence of Dirac delta distributions on the right hand side of the reformulated problem implies that \cref{eqn:C1} can only be understood in a weak sense, i.e.~if you multiply \cref{eqn:C1} by any sufficiently smooth test function and integrate from $z=0$ to $H$ the result should be true.
The homogeneous boundary conditions that appear in \cref{eqn:C1} are used in the construction of the Green's function, and guarantee that the solution will satisfy homogeneous Neumann boundary conditions whenever $b^\pm=0$.

\section*{References}


\end{document}